\documentclass[a4paper,12pt]{article}

\newcommand{\sect}[1]{\setcounter{equation}{0}\section{#1}}

\newcommand{\bea}{\begin{eqnarray}}
\newcommand{\eea}{\end{eqnarray}}
\newcommand{\be}{\begin{equation}}
\newcommand{\ee}{\end{equation}}
\newcommand{\vs}[1]{\vspace{#1 mm}}

\renewcommand{\a}{\alpha}

\renewcommand{\d}{\delta}
\newcommand{\e}{\epsilon}
\newcommand{\al}{\alpha'_{\rm eff}}
\newcommand{\alt}{\tilde{\alpha}'_{\rm eff}}
\newcommand{\dsl}{\pa \kern-0.5em /}

\newcommand{\ncos}{NCOS$_6$}
\newcommand{\go}{G_o^2}
\newcommand{\gone}{G_{o(1)}^2}
\newcommand{\gonef}{G_{o(1)}^4}
\newcommand{\half}{\frac{1}{2}}
\newcommand{\pa}{\partial}
\renewcommand{\t}{\theta}

\newcommand{\nn}{\nonumber\\}

\begin{document}
\topmargin 0pt
\oddsidemargin 0mm

%\renewcommand{\thefootnote}{\fnsymbol{footnote}}
%\begin{titlepage}
\begin{flushright}
hep-th/0210072\\
%SINP-TNP/02-? 
\end{flushright}

\vs{2}
\begin{center}
{\Large \bf  
 More on Penrose limits and non-local theories}
\vs{10}

{\large Somdatta Bhattacharya and S. Roy}
\vspace{5mm}

{\em 
 Saha Institute of Nuclear Physics,
 1/AF Bidhannagar, Calcutta-700 064, India\\
E-Mails: som, roy@theory.saha.ernet.in\\}
\end{center}

\vs{5}
\centerline{{\bf{Abstract}}}
\vs{5}
\begin{small}
We obtain the Penrose limit of six dimensional Non-Commutative Open
String (NCOS$_6$) theory and show that in the neighborhood of a particular
null geodesic it leads to an exactly solvable string theory (unlike their
counterparts in four or in other dimensions). We describe the phase
structure of this theory and discuss the Penrose limit in different phases
including Open D-string (OD1) theory. We compute the string spectrum and
discuss their relations with the states of various theories at different
phases. We also consider the case of general null geodesic for which the
Penrose limit leads to string theory in the time dependent pp-wave background 
and comment on the renormalization group flow in the dual theory.
\end{small}

\sect{Introduction}

It is well-known that string theory in the pp-wave background can be 
obtained \cite{bfhp,bfhpa} for type IIB theory by taking the Penrose limit 
\cite{pen} on 
AdS$_5\times$S$^5$ and has many interesting features. The most
notable one is that the corresponding Green-Schwarz (GS) action in the 
light-cone
gauge is exactly solvable \cite{met,mt,rt}. So one can quantize 
the theory and obtain the 
string spectrum in a straightforward way. By the Maldacena conjecture 
\cite{mal,agmoo,gkp,witt} taking
a Penrose limit (on the string/gravity side) amounts to going to a particular
subsector of the ${\cal N} = 4$, $SU(N)$ super Yang-Mills theory where both the
conformal dimension $\Delta$ and the $U(1)$ R-charge $J$ of the gauge theory
operators scales as $\Delta$, $J\,\,\sim\,\sqrt{N}$ as $N \to \infty$, keeping
$g_{YM}^2$ and $\Delta - J$ fixed. Thus using the BMN \cite{bmn} 
conjecture of exact 
correspondence between string theory in the pp-wave background and the 
subsector of gauge theory one can compute the anomalous dimensions of the
gauge theory operators from the exact string spectrum and this has 
been generalized
to many other AdS/CFT-like examples.   

The consequence of taking Penrose limits and their implications in the
subsector of dual nonconformal theories have been discussed in refs.
\cite{ps,gps,chkw,bjlm,fis,go,kp,hrv}. In
most cases Penrose limits on the gravity duals of these theories lead to
string theories in time dependent pp-wave backgrounds. For a large class of
such backgrounds, the equations of motion of the GS action can be solved
exactly, but the quantization for such systems and the construction of
states are still not well-understood \cite{fis,gps}. However, 
there is an intriguing 
connection between the associated time dependent quantum mechanical problem
and the RG flow in the dual gauge theory \cite{gps}. On the other hand, 
it has been
noticed that in six dimensions there exists a special null geodesic, for the
gravity dual of both local \cite{os} (ordinary YM or OYM) and non-local 
(Little
String Theory\footnote{The existence of this theory has been argued in
\cite{brs,sei,abks}.} (LST) \cite{hrv}, noncommutative 
YM\footnote{The gravity dual
of 4-dimensional NCYM theory has been obtained in \cite{hashi,mr} and in other
dimensions in \cite{aos,lurone}.} (NCYM) \cite{br} and open 
D$p$-branes\footnote{The
existence of OD$p$ theories have been shown in \cite{gmss} (see also
\cite{harone}) and their
supergravity descriptions are given in \cite{alior,mitro}.} (OD$p$) \cite{ka})
theories, in the neighborhood of which the Penrose limits lead to 
string theories
in time independent pp-wave backgrounds. These are very similar to the
maximally supersymmetric pp-wave limit of AdS$_5\times$S$^5$. Since string 
theories in these cases are exactly solvable, it is straightforward to obtain
the string spectrum and extract information about the states of the above
mentioned theories in the subsector corresponding to the Penrose limit. Such
discussions can be found for LST in \cite{hrv}, for 6-dimensional NCYM in 
\cite{br} and for OD5 in \cite{os}.

In this paper we study the Penrose limit of the gravity dual of another
class of non-local theories, namely, the noncommutative open string theory
\cite{sst,gmms}
in six dimensions (NCOS$_6$) \cite{har}. The supergravity configuration 
is given by the
(F,D5) bound state \cite{lur} of type IIB string theory in the 
so-called NCOS limit.
We describe the phase structure of this theory and show how at various 
energies the theory is described by OYM$_6$, LST, NCOS$_6$ and OD1 theories.
We obtain the Penrose limit of NCOS$_6$ theory in the neighborhood of a 
particular null geodesic by defining a scaling parameter in terms of the
known parameters of the theory. We show that the Penrose limit in this case
leads to an exactly solvable string theory in a time independent pp-wave
background unlike the case of four or other dimensional NCOS theories. 
Here we find that the two of the eight bosonic coordinates of the
string theory are massive and it contains both NSNS and RR three-form field
strengths. We also discuss Penrose limits at different phases of this theory
and as is known they all lead to solvable string theories. We study the
quantization of the bosonic sector of the gravity dual of NCOS$_6$ theory
in the Penrose limit, obtain the string spectrum and discuss their relations
to the states in NCOS$_6$ theory in a particular subsector. Similar discussions
are given for the different phases of this theory including the OD1 theory.
Finally we obtain the Penrose limit for a more general null geodesic of both
NCOS$_6$ theory and the OD1 theory. In this case we obtain string theories 
in 
time dependent pp-wave backgrounds which are not solvable. We briefly comment
on the RG flow in the dual theory.

The organization of this paper is as follows. In section 2, we give the gravity
dual description of NCOS$_6$ theory and describe its phase structure. The
Penrose limit of this theory and various other theories at different phases
for a particular null geodesic is discussed in section 3. The quantization
of the bosonic sector and the string spectrum for the NCOS$_6$ and other
theories at different phases are described in section 4. In section 5, we
discuss the Penrose limits of NCOS$_6$ and OD1 theories for the general null
geodesic and comment on the RG flow. Our conclusion is presented in section 6.

\sect{Supergravity description and the phase structure of NCOS$_6$}

The gravity dual description of NCOS$_6$ can be obtained from the (F,D5)
bound state configuration of type IIB string theory in the so-called NCOS 
limit and is described in refs.\cite{har,cmor}. The string metric, 
the dilaton and the
other gauge fields have the forms,
\bea
ds^2 &=& \epsilon (1+a^2r^2)^{\frac{1}{2}}ar\left[-(dx^0)^2 + (dx^1)^2 +
\frac{1}{1+a^2r^2}\sum_{i=2}^5(dx^i)^2 + \frac{MG_o^2\alpha'_{\rm eff}}{r^2}
(dr^2 + r^2 d\Omega_3^2)\right]\nonumber\\
e^\phi &=& G_o^2 (ar)\nonumber\\
B &=& \epsilon (ar)^2 dx^0 \wedge dx^1\nonumber\\
F^{(3)} &=& - 2 \epsilon M \al \epsilon_3\nonumber\\
A^{(4)} &=& \frac{\epsilon^2}{G_o^2(1+ a^2 r^2)} dx^2 \wedge dx^3 \wedge 
dx^4 \wedge dx^5
\eea
Here the parameter $\e$ is defined as $\e = \a'/\al$, where 
$\sqrt{\a'}$ is the string length scale and $\sqrt{\al}$ is the length scale 
of the NCOS$_6$ theory. The parameter $a^2 = \al/(MG_o^2)$, where $M$ is the
number of D5-branes and $G_o^2$ is the coupling constant of NCOS$_6$ theory.
`$r$' is the energy parameter defined by $r = \tilde{r}/(\sqrt{\e}\al)$,
`$\tilde{r}$' being the radial coordinate transverse to the (F,D5) 
world-volume. $d\Omega_3^2$ is the line element of the unit 3-sphere transverse
to (F,D5) world-volume and $\e_3$ is its volume-form. Note that in the
NCOS limit
$\a' \to 0$, with $G_o^2$ and $\al$ fixed and so, $\e$ is a small parameter.

The above supergravity description of \ncos\, theory is valid as long as the
curvature of the metric in (2.1) measured in $\a'$ unit as well as the
dilaton remain small i.e.
\bea
\a'{\cal R} &=& \frac{1}{M \go ar(1+a^2r^2)^{\half}} \ll 1\\
e^{\phi} &=& \go ar \ll 1
\eea
We will discuss the two cases (i) $ar \ll 1$ and (ii) $ar \gg 1$ separately.
For case (i), the condition (2.2) implies $ar \gg 1/(M\go)$ and so, along with
the relation $ar \ll 1$, we have
\be
\frac{1}{M\go} \ll ar \ll 1
\ee
where the curvature remains small. On the other hand the condition (2.3)
implies $ar \ll 1/{\go}$. So, if $\go \ll 1$, then we have the whole region
(2.4) where both the curvature and the dilaton remain small and we have
a valid supergravity description. However, we note that this conclusion is not
quite correct. The reason is, in the region $ar \ll 1$ we expect the \ncos\,
theory to reduce to OYM$_6$ theory i.e. the supergravity configuration (2.1)
should reduce to simple D5-brane configuration in the OYM-limit. It is easy
to check that this would happen only if we set along with $ar \ll 1$ and $M 
\to \infty$, $\go \to \infty$, $\al \to 0$, such that $\go \al = g_{YM}^2/(
2\pi)^3$. In this limit, (2.1) reduce to 
\bea
ds^2 &=& \a'\left[\frac{(2\pi)^{3/2}r}{\sqrt{M}g_{YM}}\left(-(dx^0)^2 + 
\sum_{i=1}^5(dx^i)^2\right) + \frac{g_{YM}\sqrt{M}r}{(2\pi)^{3/2}}
(\frac{dr^2}{r^2} + d\Omega_3^2)\right]\nonumber\\
e^\phi &=& \frac{g_{YM}r}{(2\pi)^{3/2}\sqrt{M}}\nonumber\\
F^{(3)} &=& - 2 \a' M \epsilon_3
\eea
with $B$ and $A^{(4)}$ vanishing. This is exactly the D5-brane configuration 
in the OYM$_6$ limit \cite{imsy}. So, when $ar \ll 1$, we take 
$\go \gg 1$. Therefore,
the supergravity description remains valid not in the region (2.4) but
\be
\frac{1}{M\go} \ll ar \ll \frac{1}{\go}
\ee
Note that since we are in the OYM$_6$ region, $a$ and $\go$ do not have any
obvious meaning. Substituting $a^2 = \al/(M\go)$ in (2.5) we find the range
in this case as \cite{imsy},
\be
\frac{1}{\sqrt{M}} \ll r \ll \sqrt{M}
\ee
When $ar \gg 1/{\go}$, the dilaton becomes large and we have to 
go to the S-dual
frame in order to have a valid supergravity description. In this case the
D5-brane
configuration will go over to the NS5-brane configuration by S-duality and 
OYM$_6$ limit will become the LST limit. Therefore, we will have the LST
description in the region
\be
\frac{1}{\go} \ll ar \ll 1
\ee
We will discuss how to obtain the LST supergravity description from OD1 theory 
later.

Now consider the case (ii) where $ar \gg 1$. Here we will assume $\go \ll 1$ 
and $\al\,\,=$ fixed for the weakly coupled \ncos\, theory. Note that the
curvarture condition (2.2) in this case implies $ar \gg 
1/(M \go)^{\half}$. Since for large $M$, $1/(M \go)^{\half} \ll 1$,
the curvature condition is always satisfied. Also, the dilaton condition
(2.3) implies $ar \ll 1/{\go}$ and so, combining with $ar \gg 1$, we have
the following range of $ar$ for which \ncos\, supergravity description 
remains valid,
\be
1 \ll ar \ll \frac{1}{\go}
\ee
But when $ar \gg 1/{\go}$, the dilaton becomes large and we have to go to the
S-dual frame to have a valid supergravity description. Under S-duality the
(F,D5) supergravity configuration becomes the (D1,NS5) supergravity
configuration and the \ncos\, limit goes over to the OD1 limit. To obtain the
supergravity dual of OD1 theory we make an S-duality transformation in (2.1)
which gives \cite{cmor},
\bea
ds^2 &=& \e \gone (1+a^2r^2)^{\frac{1}{2}}\left[-(dx^0)^2 + (dx^1)^2 +
\frac{1}{1+a^2r^2}\sum_{i=2}^5(dx^i)^2 + \frac{N\alt}{r^2}
(dr^2 + r^2 d\Omega_3^2)\right]\nonumber\\
e^\phi &=& \frac{\gone}{ar}\nonumber\\
dB &=& 2\e N \gone \alt \e_3\nonumber\\
A^{(2)} &=& \e (ar)^2 dx^0 \wedge dx^1\nonumber\\
A^{(4)} &=& \frac{\e^2 \gone}{1+ a^2 r^2} dx^2 \wedge dx^3 \wedge 
dx^4 \wedge dx^5
\eea
Note that in writing (2.9) we have made use of the S-duality relations among
the parameters of the two theories as \cite{gmss,cmor},
\be
\go = \frac{1}{\gone}, \qquad {\rm and} \qquad \go \al = \alt
\ee
where $\gone$ is the coupling constant and $\sqrt{\alt}$ is the length
scale in OD1 theory. Also, $N$ is the number of NS5-branes and is equal to
$M$. The parameter $a^2$ in \ncos\, theory should be written in terms of the
parameters of OD1 theory i.e.
\be
a^2 = \frac{\al}{M\go} = \frac{\gonef \alt}{N}
\ee
Note that when $ar \ll 1$, $\gone \to 0$ and $\alt\,\,=\,\,g_{YM}^2
/(2\pi)^3\,\,=$ fixed and this is precisely the LST limit. The supergravity
configuration in this limit becomes \cite{imsy}
\bea
ds^2 &=& -(dx^0)^2 + 
\sum_{i=1}^5(dx^i)^2 + \alt N
(\frac{dr^2}{r^2} + d\Omega_3^2)\nonumber\\
e^\phi &=& \frac{(2\pi)^{3/2}\sqrt{N}}{g_{YM}r}\nonumber\\
dB &=& 2 N \alt \epsilon_3
\eea
with the other fields vanishing. This is exactly the supergravity 
configuration of LST. Note that in writing (2.12) we have multiplied by 
the $g_s^{-1} = 1/{\e \gone}$ of OD1 theory (with $g_s$, the string coupling
constant), both the metric and $dB$ in (2.9) so that both the S-dual
metric of (NS5,D1) and the original metric of (F,D5) remain Minkowskian
\cite{cmor}.

So, to summarize, the phase structure of \ncos\, theory is as follows. We
have OYM$_6$ description in the range $1/(M\go) \ll ar \ll 1/{\go}$ and
LST description in the range $1/\go \ll ar \ll 1$. On the other hand we
have weakly coupled \ncos\, description in the range $1 \ll ar \ll 1/\go$ and
we have weakly coupled OD1 description if $ar \gg 1/\go$. In the first part
when $ar \ll 1$, $\go \gg 1$ and for the second part when $ar \gg 1$,
$\go \ll 1$.

\sect{Penrose limits}

In this section we obtain the Penrose limit of \ncos\, supergravity
description given in (2.1) for a particular null geodesic and discuss the same
for the other theories at different phases mentioned in the previous section.
To obtain the Penrose limit of (2.1) we first scale the coordinates
$x^{0,\ldots,5}$ as $x^{0,\ldots,5} \to \sqrt{M \go \al} x^{0,\ldots,5}$. Then
by defining a new variable $ar = e^U$ we write (2.1) as,
\bea
ds^2 &=& R^2(1+e^{2U})^{\frac{1}{2}}e^U\left[-(dx^0)^2 + (dx^1)^2 +
\frac{1}{1+e^{2U}}\sum_{i=2}^5(dx^i)^2 + 
dU^2 + d\Omega_3^2\right]\nonumber\\
e^\phi &=& G_o^2 e^U\nonumber\\
B &=& R^2 e^{2U} dx^0 \wedge dx^1\nonumber\\
F^{(3)} &=& - 2 \frac{R^2}{\go} \epsilon_3\nonumber\\
A^{(4)} &=& \frac{R^4}{G_o^2(1+ e^{2U})} dx^2 \wedge dx^3 \wedge 
dx^4 \wedge dx^5
\eea
We write $d\Omega_3^2 = \cos^2\theta d\psi^2 + d\theta^2 + \sin^2\t d\phi^2$ and
look at a null geodesic for the metric in (3.1) restricted to $(x^0,\,U,\,
\psi)$-plane. So, we set $x^{1,\ldots,5} = 0$, $\t = 0$ and so, the effective 
Lagrangian associated with this geodesic has the form,
\be
{\cal L} = -(1+e^{2U})^{1/2} e^U ((x^0)')^2 + (1+e^{2U})^{1/2} e^U (U')^2
+ (1+e^{2U})^{1/2} e^U (\psi')^2
\ee
where `prime' denotes the derivative with respect to the affine parameter along
the geodesic. Since the Lagrangian does not depend explicitly on $x^0$ and 
$\psi$, we get two constants of motion as,
\be
(1+e^{2U})^{1/2} e^U (x^0)' = E, \qquad (1+e^{2U})^{1/2} e^U \psi' = J
\ee
Substituting these in (3.2) and equating it to zero for the null geodesic
we get the evolution equation for $U$ as\footnote{It should be mentioned here
that the Euler-Lagrange equation for $U$ yields a second order differential
equation, since ${\cal L}$ explicitly depends on $U$. In general, the
equation of motion for $U$ is different from the null geodesic condition
given below in (3.4). However, since the form of the metric in (3.2) has
$-g_{00} = g_{\psi\psi} = g_{UU}$, it can be easily checked that the equation
of motion for $U$ is equivalent to the null geodesic codition as given below.},
\be
(1+e^{2U})^{1/2} e^U U' = \sqrt{1-l^2}
\ee
where we have defined $l = J/E$ and scaled the affine parameter by $E$. 
Eq.(3.4) denotes the one parameter family of evolution equation for which 
$l^2 \leq 1$. We will discuss the general case $l < 1$ (we take $l\geq 0$)
in section 5, and here we consider a special null geodesic for which
$l=1$. So, we have $U' = 0$ or $U =$ constant. Therefore the null geodesic
is now restricted to $(x^0,\,\psi)$-plane as in the case of maximally
supersymmetric AdS$_5\times$S$^5$. The null geodesic is given by $U = U_0
= {\rm constant}$, $x^{1,\ldots,5} = \t = 0$ and $x^0 = \psi = x^+$, where
$x^+$ is the affine parameter. We now define a set of new coordinates as,
\bea
U &\to & U_0 + (1+e^{2U_0})^{-1/4}e^{-U_0/2} x\nonumber\\
\theta &\to & (1+e^{2U_0})^{-1/4}e^{-U_0/2} z\nonumber\\
x^1 &\to & (1+e^{2U_0})^{-1/4}e^{-U_0/2} x^1\nonumber\\
x^{2,\ldots,5} &\to & (1+e^{2U_0})^{1/4}e^{-U_0/2} x^{2,\ldots,5}\nonumber\\
x^0 &\to & x^+ + (1+e^{2U_0})^{-1/2}e^{-U_0} x^-\nonumber\\
\psi &\to & x^+ - (1+e^{2U_0})^{-1/2}e^{-U_0} x^-
\eea
By further rescaling the coordinates as $x^+ \to x^+$, $x^- \to x^-/R^2$,
$x \to x/R$, $z \to z/R$, $x^{1,\ldots,5} \to x^{1,\ldots,5}/R$, $\phi
\to \phi$  and taking
$R \to \infty$\footnote{In this limit, it is clear that we are in the
neighborhood of the null geodesic just mentioned above.}, the 
supergravity configuration in (3.1) takes the form,
\bea
ds^2 &=& - 4 dx^+ dx^- - \vec{z}^2 (dx^+)^2 + \sum_{i=1}^5 (dx^i)^2 + dx^2
+ d\vec{z}^2\nn
e^\phi &=& \go e^{U_0}\nn
H &=& dB\,\,=\,\, 2 \frac{e^{U_0}}{\sqrt{1+e^{2U_0}}} dx^+ \wedge dx^1 \wedge
dx\nn
F^{(3)} &=& 2 \frac{e^{-U_0}}{\go \sqrt{1+e^{2U_0}}} dx^+ \wedge dz_1 
\wedge dz_2 \nn
F^{(5)} &=& 0
\eea
This is the Penrose limit of the \ncos\, supergravity configuration. 
In the above
$\vec{z}^2 = z_1^2 + z_2^2$ where $z_1 = z \cos\phi$ and $z_2 = z \sin\phi$.
We note from the metric in (3.6) that only two of the eight bosonic 
coordinates, namely, $(z_1,\,z_2)$ have constant masses and the rest are 
massless. This will lead to an exactly solvable string theory. We can 
introduce arbitrary masses for $z_1$ and $z_2$ by scaling $x^{\pm} \to 
\mu^{\pm 1} x^{\pm}$ and then (3.6) become,
\bea
ds^2 &=& - 4 dx^+ dx^- - \mu^2 \vec{z}^2 (dx^+)^2 + \sum_{i=1}^5 
(dx^i)^2 + dx^2 + d\vec{z}^2\nn
e^\phi &=& \go e^{U_0}\nn
H &=& 2\mu \frac{e^{U_0}}{\sqrt{1+e^{2U_0}}} dx^+ \wedge dx^1 \wedge
dx\nn
F^{(3)} &=& 2\mu \frac{e^{-U_0}}{\go \sqrt{1+e^{2U_0}}} dx^+ \wedge dz_1 
\wedge dz_2 \nn
F^{(5)} &=& 0
\eea
We would like to mention that in taking the Penrose limit we have taken the 
scaling parameter $R^2 = \e M \go \al \to \infty$. This can be achieved by
taking (a) $M \to \infty$, $\go$, $\al$ fixed or, (b) $M \to \infty$, $\go \to 
\infty$, $\al \to 0$ such that $\go \al = g_{YM}^2/(2\pi)^3 =$ fixed. For case
(a) we are in \ncos\, theory but for case (b) $a^2 = \al/(M\go) \to 0$ and
so $ar \ll 1$. This is the region, as we have seen in the previous section,
where we have OYM$_6$ supergravity description if $1/(M\go) \ll 
ar \ll 1/\go$ and LST supergravity description if $1/\go \ll ar \ll 1$. Note
that for the OYM$_6$ supergravity description in the region mentioned, 
the metric
and the dilaton remain the same as in (3.7) in the Penrose limit, but we
have to replace $\go e^{U_0} = \go a r_0 = g_{YM}r_0/((2\pi)^{3/2}\sqrt{M})$.
But looking at the NSNS and RR 3-forms we find that $H = 2\mu (ar_0) dx^+
\wedge dx^1 \wedge dx$ vanishes, but, $F^{(3)} = (2\mu/(ar_0\go))dx^+ 
\wedge dz_1 \wedge dz_2$ does not vanish since $ar_0\go \ll 1$. This is 
precisely the Penrose limit of the D5-brane (in the near horizon limit) 
discussed
in \cite{os}. For LST supergravity description in 
the region $1/\go \ll ar \ll 1$,
we have to go to the S-dual frame and we will discuss how to obtain this
from the OD1 theory later.

We have mentioned in the previous section that for $ar \gg 1$ and $\go \ll 1$,
we have \ncos\, description in the range $1 \ll ar \ll 1/\go$ whose Penrose
limit we have already discussed. But when $ar \gg 1/\go$, the dilaton becomes
large and we have to go to the S-dual description which is nothing but
the gravity dual of OD1 theory given in (2.9). We now discuss the Penrose limit
of this theory. In this case we scale the coordinates as, $x^{0,\ldots,5}
\to \sqrt{N\alt} x^{0,\ldots,5}$, define a new coordinate $e^U = ar$ as before
and define the scaling parameter as $R^2 = \e N \alt \gone$ (this is not the
same as in \ncos\, case), then (2.9) takes the form,
\bea
ds^2 &=& R^2(1+e^{2U})^{\frac{1}{2}}\left[-(dx^0)^2 + (dx^1)^2 +
\frac{1}{1+e^{2U}}\sum_{i=2}^5(dx^i)^2 + 
dU^2 + d\Omega_3^2\right]\nonumber\\
e^\phi &=& \frac{\gone}{e^U}\nonumber\\
dB &=& 2R^2 \e_3\nonumber\\
A^{(2)} &=& \frac{R^2}{\gone} e^{2U} dx^0 \wedge dx^1 \nonumber\\
A^{(4)} &=& \frac{R^4}{\gone(1+ e^{2U})} dx^2 \wedge dx^3 \wedge 
dx^4 \wedge dx^5
\eea
To obtain the Penrose limit we proceed exactly as in \ncos\, case. The 
evolution equation in this case takes the form,
\be
(1+e^{2U})^{1/2} U' = \sqrt{1-l^2}
\ee
The parameter `$l$' was defined before and the the `prime' denotes derivative
with respect to the affine parameter along the geodesic in 
$(x^0,\,U,\,\psi)$-plane. Again we notice that for $l=1$, $U'=0$ or, $U=$
constant is a solution to (3.9). The null geodesic is given by $U = U_0 =$
constant, $x^{1,\ldots,5} = \t = 0$ and $x^0 = \psi = x^+$ (the affine
parameter) and is restricted to $(x^0,\,\psi)$-plane. The set of new 
coordinates we now define are as follows,
\bea
U &\to & U_0 + (1+e^{2U_0})^{-1/4} x\nonumber\\
\theta &\to & (1+e^{2U_0})^{-1/4} z\nonumber\\
x^1 &\to & (1+e^{2U_0})^{-1/4} x^1\nonumber\\
x^{2,\ldots,5} &\to & (1+e^{2U_0})^{1/4} x^{2,\ldots,5}\nonumber\\
x^0 &\to & x^+ + (1+e^{2U_0})^{-1/2} x^-\nonumber\\
\psi &\to & x^+ - (1+e^{2U_0})^{-1/2} x^-
\eea
By further rescaling the coordinates as $x^+ \to \mu x^+$, $x^- \to 
x^-/(\mu R^2)$,
$x \to x/R$, $z \to z/R$, $x^{1,\ldots,5} \to x^{1,\ldots,5}/R$, $\phi
\to \phi$  and taking
$R \to \infty$, the configuration (3.8) takes the following form 
in the new coordinates,
\bea
ds^2 &=& - 4 dx^+ dx^- - \mu^2 \vec{z}^2 (dx^+)^2 + \sum_{i=1}^5 
(dx^i)^2 + dx^2 + d\vec{z}^2\nn
e^\phi &=& \gone e^{-U_0}\nn
dB &=& -\frac{2\mu}{\sqrt{1+e^{2U_0}}} dx^+ \wedge dz_1 \wedge
dz_2\nn
dA^{(2)} &=& \frac{2\mu e^{2U_0}}{\gone \sqrt{1+e^{2U_0}}} dx \wedge dx^+  
\wedge dx^1 \nn
F^{(5)} &=& 0
\eea
This is the Penrose limit of the gravity dual of OD1 theory and has also
been obtained in ref.\cite{ka}. In order to recover the 
Penrose limit of gravity
dual of LST from here in the region $\gone \ll ar \ll 1$, we also have
to set $\gone \to 0$ and $\alt = g_{YM}^2/(2\pi)^3 =$ fixed. The metric would
have the same form as given in (3.11). The dilaton $e^{\phi} = \gone/(ar_0)
= (2\pi)^{3/2} \sqrt{N}/(g_{YM} r_0)$, $dB = - 2\mu dx^+ \wedge dz_1 \wedge
dz_2$ which remains finite, but $dA^{(2)} = (2\mu\gone\alt/N) r_0^2 dx \wedge
dx^+ \wedge dx^1$ vanishes. This is exactly the Penrose limit of the gravity
dual of LST discussed in \cite{hrv}.

\sect{Quantization and the string spectrum}

In this section we discuss the quantization of the bosonic sector of the closed
string theory obtained in the previous section by taking the Penrose limit
of the gravity dual of \ncos\, theory as well as various other theories
at different phases. We will obtain the string spectrum and discuss their
relations to the states in various dual theories. The GS action for the
bosonic part has the form,
\be
-4\pi \alpha' S_b = \int d^2\sigma \left[\eta^{ab} G_{\mu\nu} \pa_a x^\mu
\pa_b x^\nu + \e^{ab} B_{\mu\nu} \pa_a x^\mu \pa_b x^\nu\right]
\ee
where $\eta^{ab} = {\rm diag}(-1,\,1)$ is the world-sheet metric and
$\e^{\tau\sigma} = 1$. We first rename the coordinates as $(z_1,\,z_2,\,
x^5,\,x^4,\,x^3,\,x^2,\,x^1,\,x) \equiv (z_1,\,z_2,\,z_3,\,z_4,
\,z_5,\,z_6,\,z_7,\,
z_8)$ and then for the background (3.7) the above action takes the form in
the light-cone gauge as,
\be
-4\pi\alpha' S_b = \int d\tau \int_0^{2\pi\alpha'p^+} d\sigma \left[\eta^{ab}
\pa_a z_i \pa_b z_i + \mu^2 z_k^2 + 4 \mu \frac{e^{U_0}}{\sqrt{1+e^{2U_0}}}
z_8 \pa_{\sigma} z_7\right]
\ee
where $i = 1,\ldots,8$ and $k = 1,2$. We have also used the light-cone gauge
$x^+ = \tau$. Let us define $Y = (z_7 + i z_8)/2$, then the equations of motion
following from (4.2) are,
\bea
\eta^{ab} \pa_a \pa_b z_k - \mu^2 z_k &=& 0, \qquad {\rm for} \qquad k=1,2\nn
\eta^{ab}\pa_a \pa_b z_l &=& 0, \qquad {\rm for} \qquad l=3,\ldots,6\nn
\eta^{ab}\pa_a \pa_b Y + 2i\mu \frac{e^{U_0}}{\sqrt{1+e^{2U_0}}}\pa_{\sigma}
Y &=& 0\nn
\eta^{ab}\pa_a \pa_b \bar{Y} - 2i\mu \frac{e^{U_0}}{\sqrt{1+e^{2U_0}}}
\pa_{\sigma} \bar{Y} &=& 0
\eea
We solve these equations by Fourier expanding the various coordinates as,
\bea
z_i &=& \sum_{n=0}^\infty \left[\frac{1}{\sqrt{4p^+\omega_n}}\alpha_n^i 
e^{-i\omega_n \tau + in\sigma/
(\alpha'p^+)} 
+ \frac{1}{\sqrt{4p^+ \omega_{-n}}}(\alpha_n^i)^\dag 
e^{i\omega_{-n} \tau - in\sigma/
(\alpha'p^+)}\right]\nn
Y &=& \sum_{n=0}^\infty \left[\frac{1}{\sqrt{4p^+\omega_n}}\alpha_n^+ 
e^{-i\omega_n \tau + in\sigma/
(\alpha'p^+)} 
+ \frac{1}{\sqrt{4p^+ \omega_{-n}}}(\alpha_n^-)^\dag 
e^{i\omega_{-n} \tau - in\sigma/
(\alpha'p^+)}\right]\nn
\eea
and similarly for $\bar{Y}$, where, 
\bea
\omega_n &=& \sqrt{\mu^2 + \frac{n^2}{(\a' p^+)^2}}, \qquad {\rm for}\quad
z_1, z_2\nn
&=& \frac{|n|}{\a'p^+},\qquad\qquad\qquad {\rm for} 
\quad z_3,\ldots, z_6\nn
&=& \sqrt{\frac{n^2}{(\a' p^+)^2} + 2\mu \frac{e^{U_0}}{(1+e^{2U_0})^{1/2}}
\frac{n}{\a' p^+}}, \qquad {\rm for} \quad Y, \bar{Y}
\eea
and we find that the oscillators obey the commutation relations,
\be
\left[\a_m^i,\,(\a_n^j)^\dag\right] = i \d_{mn} \d^{ij}, \quad
\left[\a_m^+,\,(\a_n^+)^\dag\right] = 
\left[\a_m^-,\,(\a_n^-)^\dag\right] = i \d_{mn}
\ee
where $i,j = 1,\ldots,6$. So, the bosonic part of the light-cone Hamiltonian
takes the form,
\be
2p^- = \sum_n \left[N_n^{(k)} \sqrt{\mu^2 + \frac{n^2}{(\a'p^+)^2}} +
N_n^{(l)} \frac{|n|}{\a'p^+} + (N_n^+ + N_{-n}^-)\sqrt{\frac{n^2}{(\a'p^+)^2}
+2\mu \frac{e^{U_0}}{(1+e^{2U_0})^{1/2}}\frac{n}{\a'p^+}}\right]
\ee
Now in order to relate the string spectrum to the states in \ncos\, theory
we write from (3.5)
\be
\frac{\pa}{\pa x^+} = \frac{\pa}{\pa x^0} + \frac{\pa}{\pa \psi},
\qquad \frac{\pa}{\pa x^-} = \frac{e^{-U_0}(1+e^{2U_0})^{-1/2}}{R^2}
\left(\frac{\pa}{\pa x^0} - \frac{\pa}{\pa \psi}\right)
\ee
In terms of the generators of the original $x^0$ (before rescaling by 
$\sqrt{M \go \al}$) we get,
\bea
\frac{2 p^-}{\mu} &=& i \frac{\pa}{\pa x^+}\,\,=\,\, \sqrt{M \go \al} E - J_1
\nn
2\mu p^+ &=& i \frac{\pa}{\pa x^-}\,\,=\,\, \frac{e^{-U_0}
(1+e^{2U_0})^{-1/2}}{R^2} (\sqrt{M \go \al} E + J_1)
\eea
where we have used $i \frac{\pa}{\pa x^0} = \sqrt{M \go \al} E$ and $ -i \frac{
\pa}{\pa \psi} = J_1$. We thus find a correspondence between the string 
spectrum and the states in the \ncos\, theory with energy and $U(1)$ R-charge,
\be
\sqrt{M \go \al} E, \,\, J_1 \sim M \go \al \to \infty, \quad {\rm with}\,\,\,
\sqrt{M \go \al} E - J_1 = {\rm fixed}
\ee
Thus the spectrum of strings in the background (3.7) is the same as the 
spectrum of \ncos\, theory in the regime (4.10). For $R^2 \to \infty$, we
have,
\be
\mu p^+ R^2 = e^{-U_0}(1+e^{2U_0})^{-1/2} J_1 \qquad \Rightarrow
\quad \a' p^+ = \frac{e^{-U_0}(1+e^{2U_0})^{-1/2} J_1}{\mu M \go}
\ee
So, the light-cone energy now takes the form,
\bea
\frac{2p^-}{\mu} &=& \sum_n \left[N_n^{(k)} \sqrt{1 + \frac{M^2 G_o^4
e^{2U_0}(1+e^{2U_0})n^2}{J_1^2}} +
N_n^{(l)} \frac{|n|M \go e^{U_0}(1+e^{2U_0})^{1/2}}{J_1}\right.\nn 
& & \qquad \left. + (N_n^+ + N_{-n}^-)\sqrt{\frac{M^2 G_o^4 e^{2U_0}
(1+e^{2U_0}) n^2}{J_1^2}
+ \frac{2 e^{2U_0}M \go n}{J_1}}\right]
\eea
There are three terms in the light-cone energy expression in (4.12). The 
first term corresponds to the two massive bosons $z_1$, $z_2$, the second term
corresponds to the four free massless bosons $z_3, \ldots, z_6$ and the third
term corresponds to the two complex interacting bosons $Y$, $\bar{Y}$. The
massive bosons when written in terms of two complex massive bosons $z =
(z_1 + i z_2)/2$ and $\bar{z} = (z_1 - i z_2)/2$, will carry a $U(1)_2$-charge
corresponding to the angular coordinate $\phi$, while the rest of the bosons
are $U(1)_2$-charge neutral. Actually the gravity dual of \ncos\, theory before
taking the Penrose limit has $SO(4) \simeq SU(2)_L \times SU(2)_R$ isometry
of $S^3$ and $U(1)_1 \times U(1)_2$ is the subgroup of this group corresponding
to the isometries of $\psi$ and $\phi$ of $d\Omega_3^2$. The first one
corresponds to the R-charge $J_1$ introduced earlier. Also, we note that the
\ncos\, theory does not have the full 6-dimensional Poincare invariance
because of the presence of the electric field along $z_7$-direction which is
proportional to $z_8$ (or vice-versa) as can be seen from (3.7). Thus we
have another $U(1)_3$-charge carried by the bosonic fields $Y$ and $\bar{Y}$
and the other bosons are neutral under this charge. This is the reason we 
have a further splitting of the light-cone energy (the last term in (4.12))
for the bosons $Y$ and $\bar{Y}$. We would like to point out that such 
a splitting did not happen for the case of 6-dimensional NCYM theory studied
in \cite{br} and the effect of magnetic field there was 
unobservable in the spectrum.
The reason might be that we obtained the spectrum only for the closed string
sector and the effect might be observable in the open string sector. However,
in this case, we see the effect of electric field even in the closed string
sector.

We have seen that if $ar$ lies between $1/(M\go) \ll ar \ll 1/\go$, then
the Penrose limit of \ncos\, theory (3.7) reduces to that of OYM$_6$ theory
and the metric has the same form as given in (3.7), the dilaton is given
as $e^\phi = g_{YM} r_0/((2\pi)^{3/2} \sqrt{M})$, while $dB = 0$. The RR
3-form is non-vanishing. So, the GS action for the bosonic sector will
have the same form as given in (4.2) without the last term. The bosonic
part of the light-cone Hamiltonian will then be given as,
\be
2p^- = \sum_n \left[N_n^{(k)} \sqrt{\mu^2 + \frac{n^2}{(\a'p^+)^2}} +
N_n^{(l)} \frac{|n|}{\a'p^+}\right]
\ee
where $k=1,2$ correspond to the massive bosons $z_1$, $z_2$ and $l = 3,
\ldots,8$ correspond to the rest of the free massless bosons. 
To relate the string 
spectrum with the states of OYM$_6$ theory we first find
\bea
\frac{2 p^-}{\mu} &=& i \frac{\pa}{\pa x^+}\,\,=\,\, i \frac{\pa}{\pa x^0}
+ i \frac{\pa}{\pa \psi}\,\, =\,\, \frac{\sqrt{M} g_{YM}}{(2\pi)^{3/2}}
E - J_1
\nn
2\mu p^+ &=& i \frac{\pa}{\pa x^-}\,\,=\,\, \frac{e^{-U_0}}{R^2}
(i\frac{\pa}{\pa x^0} - i \frac{\pa}{\pa \psi})\,\, =\,\,\frac{e^{-U_0}
}{R^2} (\frac{\sqrt{M} g_{YM}}{(2\pi)^{3/2}} E + J_1)
\eea
The states of OYM$_6$ theory have energy and $U(1)$ R-charge
\be
\frac{\sqrt{M} g_{YM}}{(2\pi)^{3/2}} E, \,\, J_1 \sim \frac{M g_{YM}^2}
{(2\pi)^3} \to \infty, \quad {\rm with}\,\,\,
\frac{\sqrt{M} g_{YM}}{(2\pi)^{3/2}} E - J_1 = {\rm fixed}
\ee
Taking $R^2 \to \infty$ in (4.14), the light-cone energy takes the form,
\be
\frac{2p^-}{\mu} = \sum_n \left[N_n^{(k)} \sqrt{1 + \frac{M g_{YM}^2}
{(2\pi)^3}\frac{e^{2U_0}n^2}{J_1^2}} +
N_n^{(l)} \frac{|n|\sqrt{M} g_{YM}}{(2\pi)^{3/2} J_1}e^{U_0}\right]
\ee
Here $e^{U_0} = r_0$. A similar form of light-cone energy has also been
obtained in \cite{os}.

We next discuss the spectrum for the OD1 theory in the Penrose limit by
looking at the configuration given in (3.11). The light-cone GS action
in this case takes the form,
\be
-4\pi\alpha' S_b = \int d\tau \int_0^{2\pi\alpha'p^+} d\sigma \left[\eta^{ab}
\pa_a z_i \pa_b z_i + \mu^2 z_k^2 - \frac{2 \mu}{\sqrt{1+e^{2U_0}}}
z_2 \pa_{\sigma} z_1\right]
\ee
where $i = 1,\ldots,8$ and $k = 1,2$. Defining $z=(z_1 + i z_2)/2$, the 
equations of motion following from (4.17) have the forms,
\bea
\eta^{ab}\pa_a \pa_b z_l &=& 0, \qquad {\rm for} \qquad l=3,\ldots,8\nn
\eta^{ab}\pa_a \pa_b z - \mu^2 z + \frac{2i\mu}{\sqrt{1+e^{2U_0}}}\pa_{\sigma}
z &=& 0\nn
\eta^{ab}\pa_a \pa_b \bar{z} - \mu^2 \bar{z} - \frac{2i\mu}{\sqrt{1+e^{2U_0}}}
\pa_{\sigma} \bar{z} &=& 0
\eea
We can solve these equations by Fourier expanding the coordinates 
$z_{3,\ldots,8}$ as the first expression in (4.4) and $z$, $\bar{z}$ as the 
second expression in (4.4), where now $\omega_n$ will be of the forms,
\bea
\omega_n &=& \frac{|n|}{\a'p^+},\qquad\qquad\qquad {\rm for} 
\quad z_3,\ldots, z_8\nn
&=& \sqrt{\mu^2 + \frac{n^2}{(\a' p^+)^2} + \frac{2\mu}{(1+e^{2U_0})^{1/2}}
\frac{n}{\a' p^+}}, \qquad {\rm for} \quad z, \bar{z}
\eea
and the oscillators satisfy the commutation relations,
\be
\left[\a_m^i,\,(\a_n^j)^\dag\right] = i \d_{mn} \d^{ij}, \quad
\left[\a_m^+,\,(\a_n^+)^\dag\right] = 
\left[\a_m^-,\,(\a_n^-)^\dag\right] = i \d_{mn}
\ee
where $i,j = 3,\ldots,6$. So, the bosonic part of the light-cone Hamiltonian
would be given as,
\be
2p^- = \sum_n \left[N_n^{(l)}\frac{|n|}{\a' p^+} +
 (N_n^+ + N_{-n}^-)\sqrt{\mu^2 + \frac{n^2}{(\a'p^+)^2}
+ \frac{2\mu}{(1+e^{2U_0})^{1/2}}\frac{n}{\a'p^+}}\right]
\ee
To relate the string spectrum with those of the OD1 theory we first find
\bea
\frac{2 p^-}{\mu} &=& i \frac{\pa}{\pa x^+}\,\,=\,\, i \frac{\pa}{\pa x^0}
+ i \frac{\pa}{\pa \psi}\,\, =\,\, \sqrt{N \alt} 
E - J_1
\nn
2\mu p^+ &=& i \frac{\pa}{\pa x^-}\,\,=\,\, \frac{(1 + e^{2U_0})^{-\half}}{R^2}
(i\frac{\pa}{\pa x^0} - i \frac{\pa}{\pa \psi})\,\, =\,\,\frac{(1 +
e^{2U_0})^{-\half}
}{R^2} (\sqrt{N \alt} E + J_1)\nn
\eea
where we have used $ i \frac{\pa}{\pa x^0} = \sqrt{N \alt} E$ and $ -i
\frac{\pa}{\pa \psi} = J_1$. The corresponding states in OD1 theory will have
energy and $U(1)$ R-charge
\be
\sqrt{N \alt} E\,\,\,\,\, {\rm and} \,\,\, J_1 \sim N \alt \to \infty,
\quad {\rm with}\,\,\, \sqrt{N \alt} E - J_1 = {\rm fixed}
\ee
Thus the spectrum of strings in the background (3.11) is the same as those
of the OD1 theory in the regime (4.21). For $R^2 \to \infty$, we find
\be
\mu p^+ R^2 = (1 + e^{2U_0})^{-1/2} J_1, \quad {\rm or} \quad 
\a' p^+ = \frac{(1+e^{2U_0})^{-1/2} J_1}{\mu N}
\ee
The light-cone energy therefore takes the form,
\be
\frac{2p^-}{\mu} = \sum_n \left[N_n^{(l)} \frac{(1+e^{2U_0})^{1/2} N}{J_1}
|n| + (N_n^+ + N_{-n}^-)\sqrt{1 +\frac{(1+ e^{2U_0})N^2}{J_1^2} n^2 +
\frac{2 N n}{J_1}}\right]
\ee
We have seen that in the regime $\gone \ll ar \ll 1$, the Penrose limit of 
the OD1
supergravity description reduces to that of the LST supergravity description. 
This
is described after eq.(3.11). The spectrum of LST from the quantization of
GS action has already been discussed in ref.\cite{hrv} and we will not 
repeat it
here.

\sect{Penrose limits for general null geodesics}

In the previous sections we have discussed the Penrose limit of \ncos\, 
supergravity description and various other theories at different phases
for a particular null geodesic corresponding to the parameter $l=1$. In
this section we extend it for $l<1$ and discuss Penrose limits
of \ncos\, and OD1 theories. The gravity dual of \ncos\, theory is given in
(3.1). By restricting the null geodesic in $(x^0,\,U,\,\psi)$-plane we
obtained the evolution equation for $U$ in (3.4) for the general value of
the parameter $l$. Eq.(3.4) can be solved as,
\be
\half e^U \sqrt{1+e^{2U}} + \sinh^{-1} e^U = \sqrt{1-l^2} u
\ee
where $u$ is the affine parameter along the null geodesic. One can use
this relation to formally express $e^U$ as a function of $u$ and let
us call that function as $g$, i.e., $e^U = g(u)$ which satisfies (5.1).
Now we make a coordinate change from $(x^0,\,U,\,\psi) \to (u,\,v,\,x)$
by the relations,
\bea
dU &=& \frac{\sqrt{1-l^2}}{g\sqrt{1+g^2}} du\nn
dx^0 &=& \frac{1}{g\sqrt{1+g^2}}du + 2dv + ldx\nn
d\psi &=& \frac{l}{g\sqrt{1+g^2}}du + dx
\eea
By further rescaling the coordinates $u \to u$, $v \to v/R^2$, $\t \to
z/R$, $x \to x/R$, $x^{1,\ldots,5} \to x^{1,\ldots,5}/R$, $\phi \to \phi$
and taking $R \to \infty$, the metric in (3.1) reduces to
\bea
ds^2 &=& - 4 du dv - \frac{l^2}{g\sqrt{1+g^2}} \vec{z}^2 du^2 + g \sqrt{1+g^2}
(1-l^2) dx^2 + g\sqrt{1+g^2} d\vec{z}^2\nn
& & \qquad + g\sqrt{1+g^2} (dx^1)^2 + \frac{g}{\sqrt{1+g^2}}\sum_{i=2}^5
(dx^i)^2
\eea
This is the form of the metric in `Rosen' coordinates\footnote{By `$\,\,\,\,$'
we mean that strictly speaking the metric in Rosen coordinates should have
$g_{uu} = 0$ (see for example, \cite{blafp} and references therein) which 
is not the case here  and also later in (5.9). However,
that does not prevent us to obtain the desired Brinkman form of the metric in
(5.5). The reason is, since $u$ is merely renamed by $x^+$ to go to the
Brinkman form, this term just shifts the mass$^2$ of the coordinates $\vec{z}$
as given in (5.6).}. To write it in
Brinkman form we define a new set of coordinates as,
\bea
u &\to& x^+\nn
x^1 &\to& \frac{1}{\sqrt{g}(1+g^2)^{1/4}} x^1\nn
x^{2,\ldots,5} &\to & \frac{(1+g^2)^{1/4}}{\sqrt{g}} x^{2,\ldots,5}\nn
\vec{z} &\to& \frac{1}{\sqrt{g}(1+g^2)^{1/4}} \vec{z}\nn
x &\to& \frac{1}{\sqrt{1-l^2} \sqrt{g}(1+g^2)^{1/4}} x\nn
v &\to & x^- - \frac{1}{8} \left[\frac{(g\sqrt{1+g^2})'}{g\sqrt{1+g^2}}
(x^2 + \vec{z}^2 + (x^1)^2) + \frac{(\frac{g}{\sqrt{1+g^2}})'}{(\frac{g}{\sqrt
{1+g^2}})} \sum_{i=2}^5 (x^i)^2\right]
\eea
Then the metric takes the form,
\bea
ds^2 &=& - 4 dx^+ dx^- - \left(m_z^2 \vec{z}^2 + m_x^2(x^2 + (x^1)^2) + 
m_{2,\ldots,5}^2 \sum_{i=2}^5 (x^i)^2\right) (dx^+)^2
\nn
& & \qquad + (dx)^2 + d\vec{z}^2 + \sum_{i=1}^5
(dx^i)^2
\eea
Where the mass$^2$'s associated with various coordinates are given as,
\bea
m_z^2 &=& \frac{l^2}{g^2(1+g^2)} - \frac{[\sqrt{g}(1+g^2)^{1/4}]''}{\sqrt{g}
(1+g^2)^{1/4}}\nn
&=& \frac{l^2}{e^{2U} (1+e^{2U})} + \frac{(1-l^2)(1+4e^{4U})}
{4e^{2U}(1+e^{2U})^3}\nn
m_x^2 &=&  - \frac{[\sqrt{g}(1+g^2)^{1/4}]''}{\sqrt{g}
(1+g^2)^{1/4}}\nn
&=&  \frac{(1-l^2)(1+4e^{4U})}
{4e^{2U}(1+e^{2U})^3}\nn
m_{2,\ldots,5}^2 &=&  
 - \frac{[\sqrt{g}/(1+g^2)^{1/4}]''}{\sqrt{g}/
(1+g^2)^{1/4}}\nn
&=&  \frac{(1-l^2)}
{4e^{2U}(1+e^{2U})^3} (1+8 e^{2U})
\eea
where `prime' in both (5.4) and (5.6) represents derivative with respect to
the affine parameter $u=x^+$ along the geodesic. The second line in each of
the mass$^2$ expressions in (5.6) are obtained by using the evolution equation
(3.4). The important thing to note here is that all the mass$^2$ expressions
are positive for $l<1$ at all energies as opposed to some cases noted in the
literature. However, since they are time dependent, it is not clear how
to quantize and obtain the spectrum for the associated string theory. In this
sense, for these general null geodesics the Penrose limit does not lead to
a solvable string theory.

Now we discuss the Penrose limit of OD1 supergravity description for the
general null geodesic. The supergravity configuration for OD1 theory is
given in (3.8) and the evolution equation for $U$ is given in (3.9). The
solution of the equation has the form,
\be
\sqrt{1+e^{2U}} + \frac{1}{2} \ln\left[\frac{\sqrt{1+e^{2U}} -1}
{\sqrt{1+e^{2U}}
+1}\right] = \sqrt{1-l^2} u
\ee
Let us formally define $\sqrt{1+e^{2U}} = f(u)$ and make the following
coordinate change,
\bea
dU &=& \frac{\sqrt{1-l^2}}{f} du\nn
dx^0 &=& \frac{1}{f}du + 2dv + ldx\nn
d\psi &=& \frac{l}{f}du + dx
\eea
By further rescaling the coordinates $u \to u$, $v \to v/R^2$, $\t \to
z/R$, $x \to x/R$, $x^{1,\ldots,5} \to x^{1,\ldots,5}/R$, $\phi \to \phi$
and taking $R \to \infty$, the metric in (3.8) can be written in Rosen
coordinates as,
\bea
ds^2 &=& - 4 du dv - \frac{l^2}{f} \vec{z}^2 du^2 + f
(1-l^2) dx^2 + f d\vec{z}^2
+ f (dx^1)^2 + \frac{1}{f}\sum_{i=2}^5
(dx^i)^2
\eea
To write it in Brinkman form we define a new set of coordinates,
\bea
u &\to& x^+\nn
x^1 &\to& \frac{1}{\sqrt{f}} x^1\nn
x^{2,\ldots,5} &\to & \sqrt{f} x^{2,\ldots,5}\nn
\vec{z} &\to& \frac{1}{\sqrt{f}} \vec{z}\nn
x &\to& \frac{1}{\sqrt{1-l^2} \sqrt{f}} x\nn
v &\to & x^- - \frac{1}{8} \left[\frac{f'}{f}
(x^2 + \vec{z}^2 + (x^1)^2) + \frac{(f^{-1})'}{f^{-1}}
\sum_{i=2}^5 (x^i)^2\right]
\eea
In these new coordinates the metric (5.9) takes exactly the same form
as in the \ncos\, theory given in (5.5), but the mass$^2$ expressions for the
various coordinates are different and are given as follows,
\bea
m_z^2 &=& \frac{l^2}{f^2} - \frac{(\sqrt{f})''}{\sqrt{f}}
\nn
&=& \frac{l^2}{1+e^{2U}} + \frac{(1-l^2) e^{2U}}
{4(1+e^{2U})^3} (e^{2U} - 4)\nn
m_x^2 &=&  - \frac{(\sqrt{f})''}{\sqrt{f}
}\nn
&=&  \frac{(1-l^2)e^{2U}}
{4(1+e^{2U})^3} (e^{2U} - 4)\nn
m_{2,\ldots,5}^2 &=&  
 - \frac{(f^{-1/2})''}{f^{-1/2}}
\nn
&=&  \frac{(1-l^2)}
{4(1+e^{2U})^3} e^{2U}(4 - 3 e^{2U})
\eea
Here also `prime' denotes derivative with respect to the affine parameter
$u=x^+$. The second expressions of mass$^2$ in (5.11) are obtained by using
the evolution equation (3.9). We note here that unlike in the \ncos\, case
the mass$^2$'s are not always positive. In particular, we see from (5.11)
that in the IR, $m_x^2$ becomes negative. However, we have seen in the 
previous sections that in the low energy the proper supergravity description
should be that of \ncos\, theory and not the OD1 theory where all the 
mass$^2$ are indeed positive. The presence of negative mass$^2$ indicates
a quantum mechanical instability of the associated world-sheet theory and
so OD1 theory in that case flows by RG to \ncos\, theory where all the
mass$^2$ of the world-sheet bosonic fields become positive. However, we
would like to point out that in the UV where we know that OD1 is the proper
supergravity description, the mass$^2$ are not always positive. For example, 
we note that $m_{2,\ldots,5}^2$ becomes negative in the UV. So, the 
appearance of negative mass$^2$ can not always be avoided by an RG flow
argument. Some comments on the isuue of negative mass$^2$ has been made
in ref.\cite{gps}, but a better understanding of this is clearly needed.

\sect{Conclusion}

In this paper we have studied the Penrose limit of the gravity dual of
a class of non-local theories, namely, the NCOS theory in 6-dimensions.
We discussed the phase structure of this theory and have shown how at
various energies the theory is described by OYM$_6$, LST, NCOS$_6$ and
OD1 theories. In particular, at low energies when $ar\ll 1$ and $\go \gg 1$,
we get OYM$_6$ theory in the range $1/(M\go) \ll ar \ll 1/\go$ and LST
in the range $1/\go \ll ar \ll 1$. On the other hand, at high energies when
$ar \gg 1$, we have \ncos\, theory in the range $1 \ll ar \ll 1/\go$ and
OD1 theory for $ar \gg 1/\go$. We have obtained Penrose limits in the
neighborhood of a special null geodesic and have shown that Penrose limits
of the gravity duals of all these 6-dimensional theories lead to solvable
string theories. We would like to emphasize that this is a specialty of
6-dimensional theories only. In fact, it is easy to see that Penrose limits of
gravity duals of NCOS theories in other dimensions do not lead to solvable
string theories. We have quantized the string theories obtained this way,
constructed the light-cone Hamiltonian and discussed their relations to
the states of various theories at different phases of \ncos\, theory.
Finally, we also discussed Penrose limits of the gravity duals of both
\ncos\, and OD1 theories for the general null geodesic. In these cases Penrose 
limits yield string theories with time-dependent masses for the various
bosonic fields corresponding to the target space coordinates and so they
are not solvable (in the sense discussed in the paper). We have pointed
out the appearance of negative mass$^2$ for OD1 theory and discussed the 
RG flow by which the mass$^2$ become positive in some cases.

\end{document}